\documentclass[lettersize,journal,onecolumn]{IEEEtran}
\usepackage{algpseudocode}
\usepackage{amsmath}
\usepackage{algorithm}
\usepackage{array}
\usepackage[font=small,labelfont=small]{caption}
\usepackage[caption=false,font=footnotesize,labelfont=sf,textfont=sf]{subfig}
\usepackage{textcomp}
\usepackage{stfloats}
\usepackage{url}
\usepackage{verbatim}
\usepackage{graphicx}
\usepackage{cuted}
\usepackage{multirow}
\usepackage{tabularx}
\usepackage{enumitem}
\usepackage{booktabs}
\usepackage{adjustbox}
\usepackage{xcolor}
\usepackage{subcaption} 

\usepackage{cite}
\usepackage{textcomp}
\usepackage{hyperref}
\hypersetup{hidelinks}
\usepackage{tikz}
\newcommand{\circled}[1]{%
  \tikz[baseline=(char.base)]{
    \node[shape=circle,draw,inner sep=0.8pt] (char) {\scriptsize #1};
  }%
}
\makeatletter
\algnewcommand{\LineComment}[1]{\Statex \hskip\ALG@thistlm \textbf{\texttt{//#1}}}
\makeatother

\begin{document}
\bstctlcite{IEEEtranBSTCTL}
\title{HyGra: Accelerating Network-State Simulation for LLM Training in DCNs via Adaptive Packet-Flow Granularity}

\author{
Wenyi Wang, Zheng Wu, Yanmeng Wang, Haolin Mao, Lei Han, Weibei Fan, Gaogang Xie, Fu Xiao\textsuperscript{*}%
\thanks{Wenyi Wang, Zheng Wu, Yanmeng Wang, Haolin Mao, Lei Han, Weibei Fan and Fu Xiao are with the College of Computer, Nanjing University of Posts and Telecommunications, Nanjing 210023, China (e-mail: hiwangwy@foxmail.com; zwu@njupt.edu.cn; hiwangym@gmail.com; maohaolin93@gmail.com; hanlei@njupt.edu.cn; wbfan@njupt.edu.cn; xiaof@njupt.edu.cn).}%
\thanks{Gaogang Xie is with the Computer Network Information Center, Chinese Academy of Sciences (e-mail: xie@cnic.cn).}%
\thanks{(Corresponding author: Fu Xiao.)}%
}

\maketitle

\begin{abstract}
In recent years, large language models (LLMs) have driven substantial intelligent transformation across diverse industries. Commercial LLM training is typically performed over data center networks (DCNs) comprising hundreds to thousands of GPUs, with multiple devices collocated per node. As network scale expands, inter-node communication becomes a primary bottleneck to training efficiency. Network-state simulators therefore play a crucial role by enabling cost-effective evaluation of network configurations and parallelization strategies through faithful emulation of DCN dynamics during LLM training. However, existing simulators are constrained by an efficiency-fidelity trade-off, as packet-level simulators (PLSs) incur prohibitive runtime overhead, whereas flow-level simulators (FLSs) compromise essential modeling accuracy. In this paper, we develop \texttt{HyGra}, a hybrid-granularity network-state simulator that exploits intrinsic network dynamics in LLM training to adaptively switch simulation granularity. Specifically, \texttt{HyGra} employs packet-level simulation during non-steady phases with transient fluctuations and flow-level simulation during steady phases with periodic patterns, thereby accelerating execution while preserving high fidelity. Moreover, it requires no specialized hardware, supports single-machine deployment, and is compatible with existing simulators. Experiments based on representative commercial LLM workloads, including ChatGPT, DeepSeek, and Qwen, show that \texttt{HyGra} achieves up to 15.4× speedup under single parallelization strategy and 7.8× under hybrid parallelization strategies while maintaining high accuracy.
\end{abstract}

\section{Introduction}
\begin{figure}[t!]
     \centering
    \includegraphics[width=0.6\linewidth,keepaspectratio]{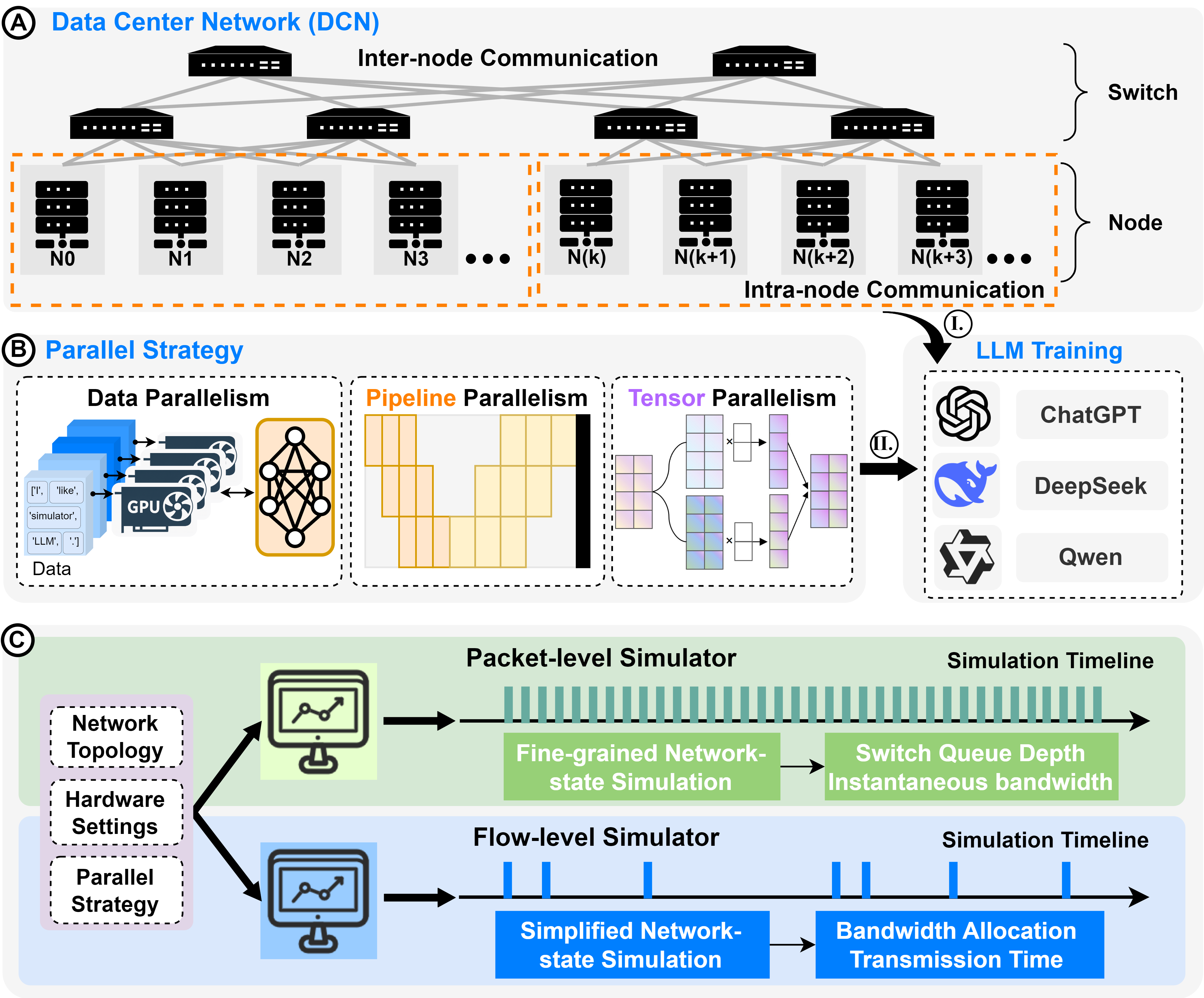}
    \caption{Simulation for LLM training in DCNs. \textcircled{\scriptsize I} refers to the hardware computational support for LLM training; and \textcircled{\scriptsize II} denotes the parallelization strategy algorithmic support.}
    \label{application}
    \vspace{-0.5cm}
\end{figure}

\IEEEPARstart{T}{he} emergence of large language models (LLMs), such as GPT\cite{gpt4}, Qwen\cite{bai2023qwen}, and LLaMA\cite{llama3}, has driven rapid advances in artificial intelligence, accelerating the intelligent transformation of a wide range of industries. Owing to the massive model sizes, the limited compute and memory capacity of a single GPU renders distributed training indispensable, with modern commercial LLMs requiring hundreds to thousands of GPUs to train collaboratively\cite{2023ashl}. In practice, these GPUs are organized within a data center network (DCN), as illustrated in Fig.\ref{application}\textcircled{\scriptsize A}, where multiple GPUs are colocated within a node, and different nodes are interconnected through multi-layer switches\cite{2024crux},\cite{2025atlahs}. Together with distributed parallel strategies-such as data parallelism (DP), pipeline parallelism (PP), and tensor parallelism (TP) shown in Fig. \ref{application}\textcircled{\scriptsize B}-the DCN constitutes the fundamental infrastructure for large-scale LLM training\cite{PP},\cite{osdi22},\cite{DP}. As DCNs have scaled from hundreds to thousands of nodes in recent years, inter-node communication has emerged as a dominant bottleneck in training latency\cite{sc21}, \cite{sc25}. Unlike intra-node communication, which typically leverages ultra-high-bandwidth interconnects such as NVLink, inter-node communication must traverse network links and multiple layers of switches, resulting in higher transmission latency\cite{sigcomm24}. Moreover, even under fixed hardware infrastructures, variations in DCN topology and parallel strategies can lead to orders-of-magnitude differences in inter-node communication latency\cite{osdi22},\cite{nsdi23}.

To mitigate inter-node communication overhead, existing approaches primarily rely on empirical tuning of system parameters such as network topology configurations and parallel strategies, guided by expert knowledge and iterative feedback from real training runs\cite{sc21}, \cite{2024megascale}. However, this experience-driven process is both time-consuming and cost-prohibitive\cite{noh2023flexblock}; for instance, training DeepSeek-V3 (67B parameters) reportedly required nearly two months and incurred costs of several million dollars\cite{deepseekv3-1}. Although analytical and model-based approaches such as TopoOpt\cite{nsdi23}, Daydream\cite{2020DaydreamAE} and Alpa\cite{osdi22}, have been proposed to estimate inter-node communication latency under different network architectures and parallel strategies, achieving accurate estimation remains challenging. This is because communication latency is jointly determined by DCN topology, hardware characteristics, and parallel strategies, leading to inevitable deviations between analytical estimates and actual network behavior\cite{2025simai}. Consequently, there is a strong need for efficient and accurate network-state simulators that can faithfully simulate DCN dynamics during LLM training, enabling cost-effective exploration of network configurations and parallel strategies to reduce inter-node communication overhead and overall training latency. 

Motivated by the potential of network-state simulators to improve efficiency and reduce costs, recent studies have developed a variety of simulation frameworks for LLM training over DCNs. From the perspective of simulation granularity, existing simulators can be categorized into packet-level simulators (PLSs) and flow-level simulators (FLSs). As illustrated in Fig. \ref{application}\textcircled{\scriptsize C}, PLS achieves fine-grained network-state simulation by explicitly modeling per-packet transmission dynamics, including instantaneous bandwidth and switch queue depth\cite{ns3},\cite{OMNET++}. In contrast, FLS abstracts packet-level behavior by estimating transmission time through bandwidth allocation algorithms that compute the average bandwidth assigned to each flow\cite{deepqueuenet}. However, PLS typically suffer from prohibitive time overheads that limit scalability, while FLS sacrifice essential modeling fidelity for execution speed\cite{li2024m3}. Moreover, some existing simulators depend on multi-node\cite{pdes} or specialized hardware environments\cite{2025spme}, which further limits their practical usability. These constraints hinder the efficiency, scalability, and applicability of current network-state simulators for large-scale LLM training in DCNs. Consequently, there is a critical need for developing a simulator that simultaneously deliver high fidelity and accelerated execution, while remaining lightweight and easy to deploy.
\subsection{Related Work}
According to modeling granularity, existing simulators are primarily single-granularity systems and can be furthur categorized into PLS and FLS. PLS simulate network states by modeling per-packet behaviors, including generation, transmission, and reception. Representative simulators include ns-3\cite{ns3} and OMNeT++\cite{OMNET++}, which focus on capturing fine-grained inter-node communication dynamics at nanosecond time scales to analyze the performance of congestion control and load balancing algorithms; while gem5\cite{binkert2011gem5} and CNSim\cite{2024CL} enhance packet-level fidelity by explicitly modeling the intra-node data transfer path from memory to the network interface. However, in large-scale LLM training scenarios, modeling every packet incurs prohibitive computational overhead, often requiring hours or even days of simulation to reproduce only seconds of network traffic\cite{li2024splitsim}. In contrast to PLS, FLS aggregates packet-level behaviors into coarse-grained flow abstractions and focuses on flow-level events, including flow arrival and flow completion. At the core of flow-level simulators is the continuous computation of bandwidth allocation, where flow completion time is directly derived by distributing bandwidth according to max-min fairness or water-filling\cite{mmf-wf}. Recent work, such as Soroush~\cite{2023MF}, extends these traditional algorithms to multipath topologies through adaptive weighting and single-pass optimization techniques. Leveraging similar flow abstractions, simulators like SimGrid\cite{simgrid} employ analytical fluid models to evaluate scheduling and resource management algorithms for distributed applications at scale. m4\cite{m4} replaces the analytical formulas of traditional flow-level simulators with the graph neural network module to directly estimate the effects of queuing and congestion. While FLS achieves orders-of-magnitude speedups, the coarse-grained abstraction limits its fidelity by omitting packet-level transient dynamics. 

A recent work, Wormhole\cite{wormhole}, accelerates simulation by exploiting traffic patterns, such as repetitive contention behaviors and steady-state characteristics. Although Wormhole achieves promising performance, the complex traffic dynamics in real-world DCNs are more likely to introduce additional estimation errors and simulation overhead. In contrast, we incorporate multiple criteria to monitor complex network states and devise an adaptive switching mechanism to accelerate simulation, thereby achieving a better trade-off between simulation accuracy and efficiency.

In order to improve the efficiency of high-precision PLS, recent studies have pursued acceleration by scaling computational resources. One line of work pursues this goal by scaling computational capacity through the addition of physical nodes. For example, pdns\cite{2004pdns}\cite{pdes} decomposes large-scale network topologies into logical processes executed in parallel across multiple physical servers; while multiverse\cite{2025spme} to offload batched tasks onto massively parallel GPUs. However, such approaches demand substantial computational infrastructure, resulting in high deployment costs. Another line of work focuses on exploiting parallelism on multi-core CPUs in single-machine. Systems such as Unison\cite{unison} and Dons\cite{dons} leverage fine-grained task partitioning and parallel scheduling to utilize multi-core processors effectively. Nevertheless, these methods typically rely on complex time synchronization and causality maintenance mechanisms, resulting in high deployment complexity. Therefore, it is necessary to design an acceleration method that is easier to deploy.

\subsection{Contribution}
In this paper, we present \texttt{HyGra}, a single-machine, hybrid-granularity network-state simulator. Unlike prior approaches that improve efficiency by scaling computational power, we subtly exploit the intrinsic characteristics of network dynamics during LLM training. Specifically, network state can be partitioned into non-steady phases marked by transient fluctuations and bursty dynamics, and steady phases characterized by periodic and homogeneous flow patterns. Motivated by this, \texttt{HyGra} adaptively simulates network state via fine-grained packet level for non-steady phases and efficient flow level for steady phases. This design accelerates LLM training simulation via packet-flow hybrid granularity, preserves high fidelity, and supports easy deployment on a single machine. Our main contributions are as follows:
\begin{itemize}
    \item \textbf{Adaptive Packet- and Flow-level Simulation:} By analyzing network traffic during LLM training, we observe that network states alternate regularly between non-steady phases with transient fluctuations and steady phases exhibiting periodic stability. Motivated by this, we apply packet-level simulation during non-steady phases and flow-level simulation during steady phases, achieving substantial acceleration on a single machine with limited computational resources while preserving high fidelity.
    \item \textbf{Development of \texttt{HyGra}:} Building on the proposed hybrid-granularity simulation framework, we develop \texttt{HyGra}, a practical single-machine hybrid-granularity network-state simulator. \texttt{HyGra} adopts a layered architecture comprising a control layer and an execution layer. The control layer monitors traffic bandwidth and queue depth to identify steady-state phases and determine packet-to-flow and flow-to-packet switching points. The execution layer dynamically adjusts simulation granularity and reconstructs the necessary network states during transitions. Besides, \texttt{HyGra} achieves seamless compatibility with existing general simulators such as ns3.
    \item \textbf{Seamless Granularity Transition:} Implementing a hybrid-granularity simulator requires addressing two key challenges. The first is how to determine the optimal time points for switching. The second is how to achieve near-lossless granularity transitions. To address these, we design a packet-to-flow switching algorithm to accurately identify granularity switching points, preventing both premature and delayed switching, and a Transformer-based queue state restoration algorithm to recover detailed network state when reverting from flow- to packet-granularity simulation.
    \item \textbf{Experiments:} We conduct a comprehensive evaluation of \texttt{HyGra} performance for simulating network state under different parallelization strategies, including PP, TP and DP. Besides, we assess its performance on several state-of-the-art LLMs, such as ChatGPT, DeepSeek and Qwen. The results show that, compared to PLS, \texttt{HyGra} achieves up to 15.4$\times$ speedup in simulation time under single parallelization strategy and up to 7.8$\times$ under complex multiple parallelization strategies. The source code is available at https://anonymous.4open.science/r/flow-input-1B8A.
\end{itemize}
\textbf{Synopsis:} Section II establishes the motivation for our work by analyzing the trade-offs in existing simulators and the characteristics of LLM training traffic. Section III introduces the overview of our proposed hybrid-granularity simulator, \texttt{HyGra}, and outlines its main challenges. Sections IV and Section V describe the design of its layered architecture. Section VI then provides a comprehensive experimental evaluation of \texttt{HyGra}. Section VII discusses the feasibility of \texttt{HyGra} for optimizing LLM training. Finally, Section VIII concludes the paper with key findings and future directions. 

\section{Motivation}
\subsection{Performance Between Packet- and Flow-Level Simulators}
The primary objective of network-state simulators for LLM training in DCNs is to capture communication dynamics, especially inter-node bandwidth and latency. The performance of such simulators is typically evaluated along two aspects:
\begin{itemize}
    \item \textbf{Efficiency:} This metric measures the computational cost of simulation, including the wall-clock runtime required to execute the whole simulation process and the associated hardware resource utilization, such as CPU utilization and memory consumption\cite{2025simai},\cite{simbricks}.
    \item \textbf{Fidelity:} This metric evaluates simulator accuracy by comparing the relative error of key statistics against high-precision packet-level simulation results, including per-flow transmission time, overall job completion time, and effective link throughput\cite{li2024m3},\cite{dons}.
\end{itemize}
\begin{figure}[t!]
   \centering
  \includegraphics[width=0.46\textwidth]{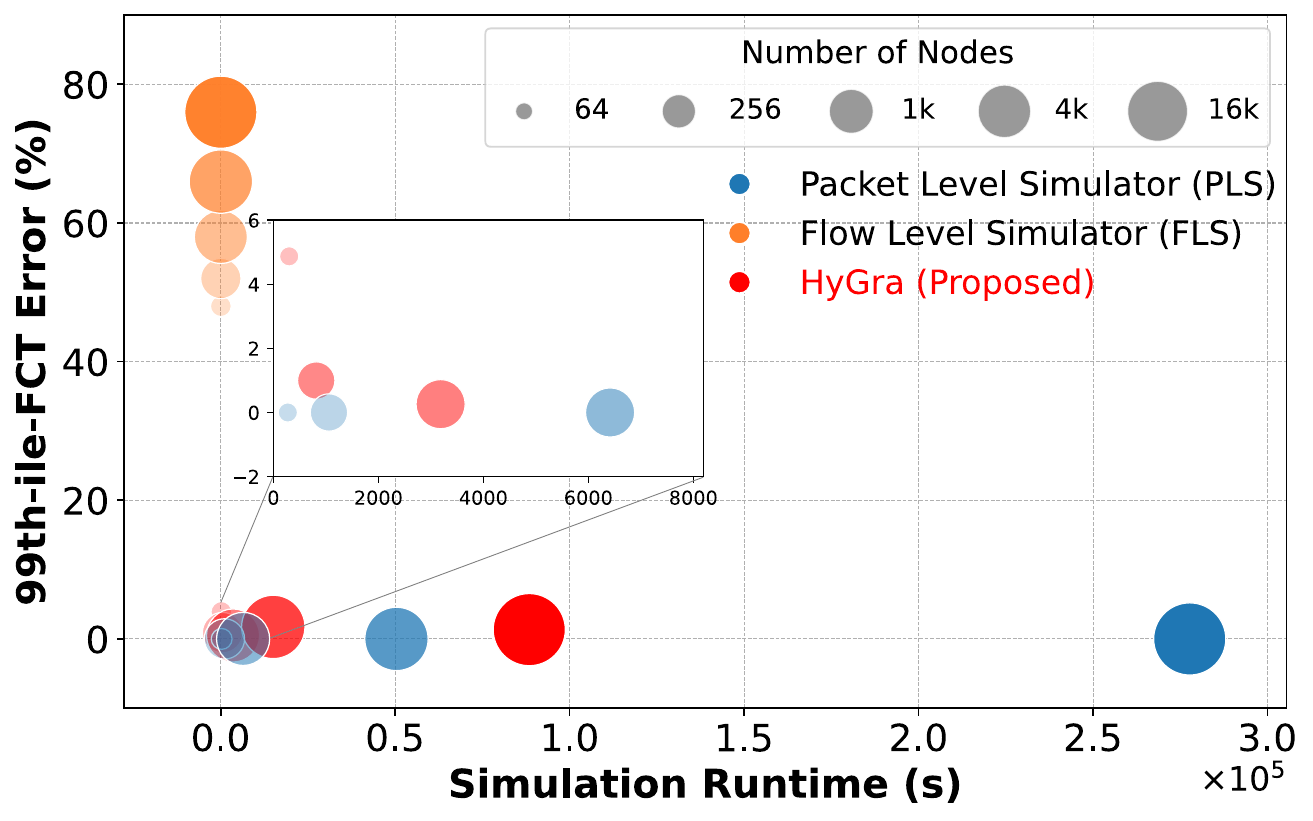}
  \caption{Performance between packet-level and flow-level simulations (DP strategy).}
  \label{Pre}
\end{figure}

\begin{figure}[t!]
  \centering
  \includegraphics[width=0.5\textwidth]{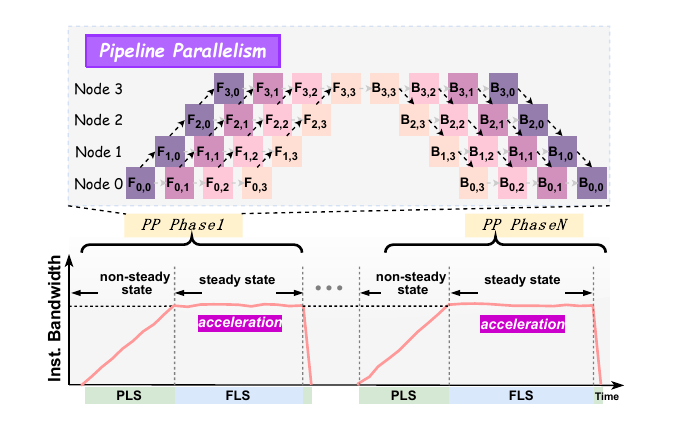}
  \caption{Instantaneous bandwidth dynamics during LLM training (PP strategy).}
  \vspace{-0.5cm}
  \label{PP_Guide}
\end{figure}
As discussed in Section I, PLS provides high-fidelity modeling but suffer from low simulation efficiency, whereas FLS achieve high execution efficiency through abstraction at the cost of reduced accuracy. As an illustrative example, Fig. \ref{Pre} quantitatively compares the simulation runtime and flow completion time (FCT) error\footnote{FCT error is defined as the relative error calculated for each individual flow in the FLS by comparing its completion time against the PLS baseline. In this paper, we specifically report the 99th percentile of these aggregated errors\cite{2010dctcp}.} of PLS and FLS under a single DP strategy with varying node scales, where simulation runtime reflects efficiency and FCT error measures simulation fidelity. The results indicate that FLS suffers from severe fidelity degradation, with FCT errors exceeding 50\%, and the error remarkably increases as the node scales. In contrast, PLS maintains high accuracy but suffers from exponential runtime growth as the node size increases. 

To further improve the efficiency of high-fidelity simulation, existing approaches follow two directions: parallel packet-level execution and analytical cost estimation. The former preserves detailed network-state evolution but still requires processing every packet event, incurring additional resource consumption and synchronization overhead. The latter improves efficiency through abstraction but cannot accurately capture queue evolution, transient flow contention, or shared-bottleneck interactions, resulting in inaccurate performance estimation under realistic DCN scenarios. \texttt{HyGra} combines the advantages of both approaches by retaining packet-level simulation during non-steady phases and adopting flow-level simulation during steady phases. This design reduces unnecessary packet processing while preserving the network-state required for accurate evaluation and remaining compatible with parallelization techniques.

\subsection{Traffic Characteristics of LLM Training}
During LLM training within DCNs, the network state dynamics generally follow a periodical behavior. This is because decentralized LLM training inherently executes ordered sequences and fixed communication schedules. Taking PP as an example, Fig. \ref{PP_Guide} illustrates the regularly changing characteristics of instantaneous inter-node bandwidth during training. Specifically, bandwidth increases gradually during the warm-up phase as sequentially injected micro-batches trigger each node to transmit hidden-layer activations to its downstream successor in a staged manner. Once the pipeline becomes fully saturated, with all nodes concurrently executing forward and backward propagation computation, bandwidth stabilizes at a steady level characterized by persistent, long-lived flows. During the cool-down phase, bandwidth rapidly declines to zero as the final micro-batches complete computation and the pipeline flushes\cite{LLMC2}. Therefore, both the warm-up and cool-down phases are considered non-steady phases. Similarly, the network state under DP and TP strategies also exhibits such regular periodical patterns, where inter-node model parameter or gradient synchronization immediately follows the forward and backward propagation computations on each GPU.

Based on the analysis above, the network state during LLM training can be divided into two phases according to key metrics including bandwidth and queue depth: 
\begin{itemize}
    \item \textbf{Non-steady Phase:} Characterized by transient fluctuations in bandwidth and queue depth, typically occurring at the beginning of training or during traffic bursts. Fine-grained modeling is required for this phase to accurately capture short-term variations and protocol dynamics.
    \item \textbf{Steady Phase:} Defined by the stabilization of bandwidth and queue depth within a bounded range, where flow behaviors become periodic and homogeneous. This phase is suitable for coarse-grained modeling and simulation acceleration.
\end{itemize}

As depicted in Fig. \ref{PP_Guide}, the above two phases alternate cyclically during LLM training. Inspired by this, the core idea of our designed simulator, \texttt{HyGra}, is to employ accurate packet-level simulation during non-steady phases and switch to flow-level simulation during steady phases. This hybrid-granularity approach ensures high fidelity while significantly reducing simulation runtime, thereby improving computational resource utilization in large-scale LLM training.
\begin{figure*}[t!]
    \centering
    \includegraphics[width=0.8\textwidth,keepaspectratio]{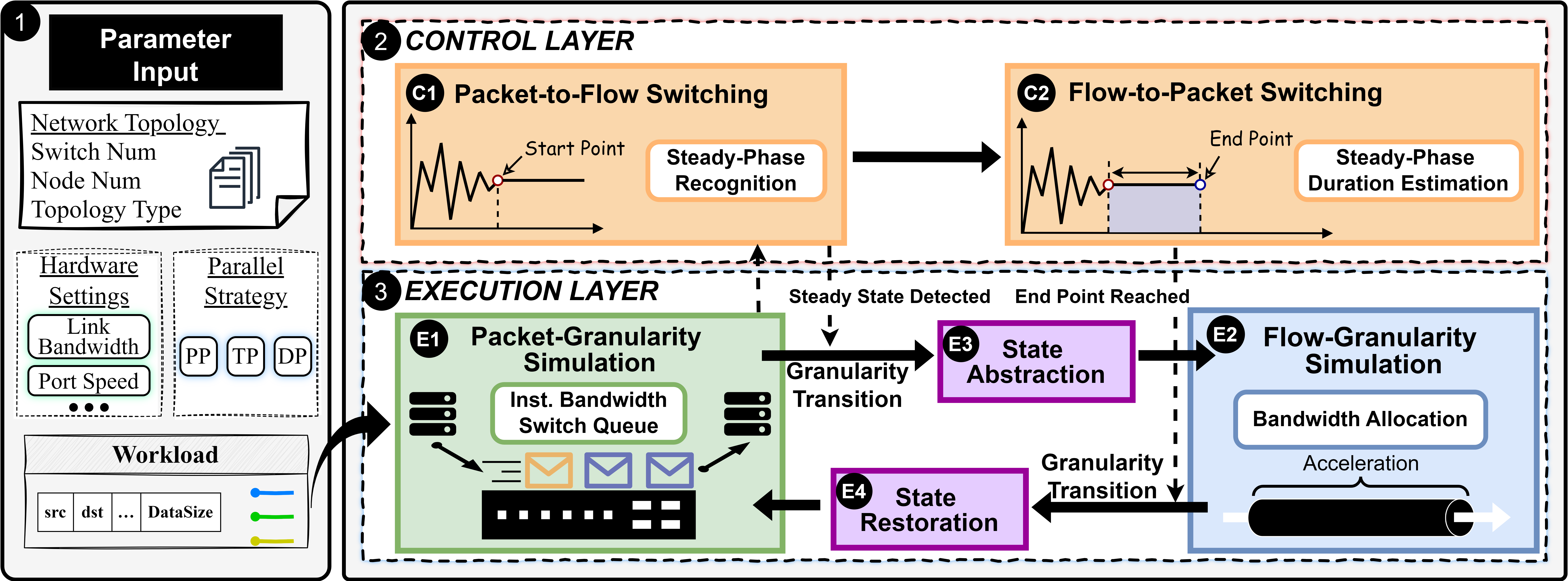}
    \caption{System Design of \texttt{HyGra}}
    \label{HyGra}
\end{figure*}

\subsection{Easy Deployment on Single Machine}
Existing approaches that accelerate PLS by expanding computational resources face two key challenges. First, reliance on specialized high-performance hardware raises deployment barriers and costs. Second, parallelization across multi-core CPUs or distributed machines has stringent requirements for time synchronization and causal consistency, significantly increasing deployment complexity. Although large-scale LLM training runs on GPU clusters, network simulation in our target scenarios is primarily used for pre-deployment design-space exploration. Therefore, requiring additional multi-node simulation environments would increase resource requirements and system complexity, limiting efficient evaluations across diverse training and network configurations. In contrast, motivated by the periodical communication pattern during LLM training analyzed in previous Section II-B, we achieve simulation acceleration through algorithmic design rather than hardware resource scaling. By leveraging hybrid-granularity simulation, our approach enables high-fidelity and efficient execution on a single commodity machine, eliminating the need for dedicated hardware and avoiding the synchronization overhead inherent in parallel architectures.

\section{System Design}
\subsection{Challenges of Hybrid Granularity-Based Simulator}
Achieving a balance between simulation efficiency and fidelity via adaptive packet-flow granularity necessitate addressing the following two challenges:
\begin{itemize}
    \item \textbf{How to determine the appropriate time point for switching simulation granularity?} The switching point directly defines the duration of the flow-level simulation. An excessively long flow-level duration may overlook critical network details, compromising accuracy, whereas a short duration results in approximately full packet-level simulation with prohibitive runtime. Therefore, it is crucial to design an efficient granularity switching algorithm to determine appropriate flow-level duration.
    \item \textbf{How to achieve near-lossless transition of simulation granularity?} Transitioning from packet to flow granularity is straightforward, as it simply collapses fine-grained packet events into coarse flow abstractions. However, the reverse transition from flow back to packet granularity is significantly more complex, as it requires restoring switch queue states that were omitted during flow-level simulation. To address this asymmetry and minimize potential cumulative errors, designing a robust state recovery mechanism is crucial.
\end{itemize}

\subsection{Overview of \texttt{HyGra}}
To address above challenges while ensuring compatibility with existing simulators, \texttt{HyGra} adopts a hierarchical architecture comprising a control layer and an execution layer as illustrated in Fig. \ref{HyGra}. This decoupled design reflects the distinct requirements of the two tasks: the control layer conducts global observation and decision making to identify granularity switching points, while the execution layer conducts network-state simulations at different granularities, as well as the transitions between them.

\subsubsection{Parameter Input}
As illustrated in Fig. \ref{HyGra}\circled{1}, the parameter initialization phase configures the simulation environment using network topology parameters such as the number of nodes and topology type, hardware settings such as link bandwidth and port speed, and parallelization strategies including PP, TP, and DP.

\subsubsection{Control Layer} 
As shown in Fig. \ref{HyGra}\circled{2}, the responsibility of this layer is to establish the temporal boundaries of flow-level simulation and thus consists of two operations:
\begin{enumerate}[label=\protect\circled{C\arabic*}]
    \item \textbf{Packet-to-Flow Switching:} This operation monitors per-flow bandwidth stability to determine when the network enters a steady phase, thereby identifying the start point for flow-level simulation.
    \item \textbf{Flow-to-Packet Switching}: This operation computes the duration of the steady phase to identify the end point of flow-level simulation.
\end{enumerate}
The detailed algorithms to achieve these switching operations are presented in the subsequent Sections IV.

\subsubsection{Execution Layer} 
As illustrated in Fig. \ref{HyGra}\circled{3}, once a granularity transition is triggered, the execution layer performs the network state transformations. It abstracts packet-level state into flow-level representations when entering flow granularity, and reconstructs packet-level state when resuming packet granularity. This layer comprises the following operation modules:
\begin{enumerate}[label=\protect\circled{E\arabic*}]
    \item \textbf{Packet-Granularity Simulation:} This module provides high-fidelity packet-level simulation for the network state during non-steady phases.
    \item \textbf{Flow-Granularity Simulation:} This module executes effficient flow-level simulation for the steady phases.
    \item \textbf{State Abstraction:} After the Packet-to-Flow Switching is triggered, this module performs the transition from packet-level to flow-level simulation via abstracting detailed packet states into flow-level parameters.
    \item \textbf{State Restoration:} Once the flow-granularity simulation terminates, this module reconstructs network state ignored during flow-level execution and resumes packet-granularity simulation.
\end{enumerate}
The detailed steps of operation modules will be discussed in Section V.

\subsubsection{\texttt{HyGra} Workflow}
The execution layer of \texttt{HyGra} follows a cyclic hybrid simulation workflow, while the control layer operates concurrently to determine granularity switching points. By default, \texttt{HyGra} begins in a non-steady phase and executes packet-level simulation (E1). Upon a packet-to-flow transition decision (C1), state abstraction is triggered (E3), and the simulator switches to flow-granularity simulation (E2) to accelerate progress. When the flow-level duration expires, a flow-to-packet transition (C2) is initiated, during which state restoration (E4) reconstructs the packet-level network state, and packet-granularity simulation (E1) resumes.

\section{Architecture of Control Layer}
As the decision-making core of \texttt{HyGra}, the control layer determines when to switch simulation granularity. As shown in Fig. \ref{HyGra}\circled{2}, the control layer identifies the start and end points of steady phase, which directly determine the duration of flow-granularity simulation, thus influencing both efficiency and fidelity of the simulator as discussed in the first challenge of Section III-A. To select appropriate switching points, \texttt{HyGra} employs a practical steady-state recognition and a duration estimation methods, both relies solely on network bandwidth and queue-length information.
\subsection{Packet-to-Flow Switching}
Determining the start point of flow-granularity simulation requires identifying when the system enters a steady phase. This assessment is based on two key metrics: bandwidth and queue depth. Bandwidth fluctuation can reflect the stability of active flows and therefore serves as the primary criterion for steady-state recognition, while queue depth can capture transient bursts or congestion, and thus serves as a secondary safeguard. The detailed assessment steps are as follows:
\begin{enumerate}
    \item \textbf{Bandwidth Stability Check: } This step performs a “per-flow” check, where each active flow is evaluated independently. At first, the instantaneous bandwidth of each flow $f_i\in \mathcal{F}_\mathrm{active}$ is sampled within a sliding window, and its variation is evaluated over consecutive windows. With denoting $\mathcal{B}_i$ as the bandwidth samples recorded for each flow $f_i$, and $\epsilon_{bw}$ representing as the steady-state threshold, if the bandwidth variation satisfies
    \begin{equation}
        v_i = \max(\mathcal{B}_i) - \min(\mathcal{B}_i) < \epsilon_{bw},
    \end{equation}
    the flow $f_i$ is classified as stable.
    \item \textbf{Queue Depth Stability Check:} This step performs a “per-port” check. Unlike the per-flow check that evaluates the bandwidth stability of individual flows, the per-port check captures the stability of switch queue states, which cannot be directly inferred from flow bandwidth. Therefore, bandwidth reflects flow transmission stability, while queue depth reflects the remaining queue variations at switch ports. Since queue depth typically converges after bandwidth becomes stable, it is evaluated only when all flows are classified as stable. The detection method is analogous to that used for bandwidth, and the system is considered to enter a steady phase only when all active flows are stable. Once the steady-state is detected, packet-to-flow switching is operated. 
\end{enumerate}
The final steady-state decision is globally determined by the stability of all active flows and monitored ports. The detailed procedure of packet-to-flow switching is summarized in Algorithm 1.
\begin{algorithm}[t!]
\caption{Packet-to-Flow Switching}
\label{algo1}
\small
\begin{algorithmic}[1]
\Require
  Active flow set $\mathcal{F}$;
  per-flow\ bandwidth sets $\{\mathcal{B}_i \mid f_i \in \mathcal{F}\}$;
  per-port\ queue sets $\{\mathcal{A}_u\}$;
  bandwidth stability threshold $\epsilon_{bw}$;
  queue stability threshold $\epsilon_{q}$;
  required stable rounds $N_\mathrm{stable}$.
\Ensure $flag = \{0,1\},$ $0$ denotes maintaining the current granularity, and $1$ indicates preparing to switch granularity.

\State $c_{\text{stable}} \gets 0$

\While{$c_{\text{stable}} < N_\mathrm{stable}$}
  \State Update all $\{\mathcal{B}_i\}$ and $\{\mathcal{A}_u\}$ with latest samples;
  \State $flag \gets 1$;
  \LineComment{Bandwidth stability check}
  \For{each $f_i \in \mathcal{F}$}
    \State Compute bandwidth variation $v_i$ by (1);
    \If{$v_i \geq \epsilon_{bw}$} $flag \gets 0;$ \textbf{break;} \EndIf
  \EndFor
  \LineComment{Queue-depth stability check (safeguard)}
  \If{$flag = 1$}
    \For{each switch port $u$}
      \State Compute queue depth variation $v_u$ similarly to (1);
      \If{$v_u \geq \epsilon_q$} $flag \gets 0;$ \textbf{break;} \EndIf
    \EndFor
  \EndIf
  \LineComment{Stable round count}
  \If{$flag = 1$}
    \State $c_{\text{stable}} \gets c_{\text{stable}} + 1$;
  \Else
    \State $c_{\text{stable}} \gets 0$;
  \EndIf
\EndWhile

\State \textbf{return} $flag$
\end{algorithmic}
\end{algorithm}

\subsection{Flow-to-Packet Switching}\label{endpoint}
The end time of flow-granularity simulation, namely the start point for resuming packet-granularity simulation, is determined by accurately estimating the duration of steady-state. This process consists of three key stages:

\textbf{1) Bandwidth Reallocation:} For all steady-state flows $f_j \in \mathcal{F}_\mathrm{steady}$, \texttt{HyGra} employs a Max-Min Fairness (MMF)-based mechanism\cite{max-min} to reallocate bandwidth. Specifically, at first the initial bandwidth assignment is given by:
\begin{equation}
    B_j^\mathrm{init} = \min(B_j^{\rm inst}, B_j^{\min}),
\end{equation}
where $B_j^{\rm inst}$ is the instantaneous bandwidth at the end of latest packet-level simulation and $B_j^{\min}$ denotes the guaranteed minimum bandwidth. Since the initial allocation of link resources may leave residual capacity, and considering that data transmissions in DCNs typically follow weighted-fair sharing, the module redistributes the remaining bandwidth among steady flows $f_j$ by:
\begin{equation}
    \begin{aligned}
        B_j^\mathrm{Re} = B_j^\mathrm{init}
        + \underbrace{\frac{C^\mathrm{total}-\sum_{k=1}^{n} B_k^{\mathrm{init}}}{\sum_{k=1}^{n} W_k^\mathrm{pri}} \cdot W_j^\mathrm{pri}}_{\overset{\triangle}{=}\Delta B_j}.
    \end{aligned}
\end{equation}
Here, $\Delta B_j$ denotes the reallocated bandwidth compensating for each steady-state flow $f_j$, $C^\mathrm{total}$ is the total available bandwidth of transmission link, and $W_j^{\mathrm{pri}}$ represents the priority weight of flow $f_j$.

\textbf{2) Steady-State Duration Estimation:} Since the completion of any steady-state flow in $\mathcal{F}_\mathrm{steady}$ will result in the network state leaving the steady phase, the steady-state duration is determined by the earliest finishing flow:
\begin{equation}
     \tau_\mathrm{steady} = \min_{f_j \in \mathcal{F_\mathrm{steady}}} \underbrace{L_j/{B_j^\mathrm{Re}}}_{\overset{\triangle}{=}\tau_j}
\end{equation}
where $L_j$ denotes the data size to be transmitted by flow $f_j$, $B_j^\mathrm{Re}$ is the reallocated bandwidth obtained by (3), and $\tau_j$ is the remaining completion time of flow $f_j$. Then, with denoting $T_\mathrm{start}$ as the start time point of the current flow-granularity simulation, the end time point is calculated by:
\begin{equation}
    T_\mathrm{end} = T_\mathrm{start} + \tau_\mathrm{steady}.
\end{equation}

\section{Architecture of Execution Layer}
The execution layer of \texttt{HyGra} conducts network-state simulation across different granularities and orchestrates the transitions between them. To simplify deployment and ensure system compatibility, we employ a established PLS to model fine-grained packet dynamics, while excuting the relatively straightforward flow-level simulation through customized FLS. Besides, since granularity transitions inevitably introduce network state errors that affect simulation fidelity, as discussed in the second challenge of Section III-A, \texttt{HyGra} incorporates a Transformer-based state restoration mechanism to achieve near-lossless granularity transition. 
\subsection{Packet- and Flow-Granularity Simulation}
\texttt{HyGra} integrates packet-level and flow-level simulation to leverage their complementary strengths. Packet-granularity simulation captures microscopic network behavior during non-steady phases, including per-packet transmission, protocol processing, and switch queue evolution. In contrast, flow-granularity simulation efficiently models macroscopic network behavior during steady phases, such as bandwidth allocation and transmission delay. This hybrid design achieves substantial simulation acceleration while maintaining high fidelity. In \texttt{HyGra}, the packet-granularity simulation is implemented upon established commercial PLS, such as ns-3\cite{ns3}, while the flow-granularity simulation is custom-designed and relies on analytical estimations of bandwidth allocation and transmission delay, as defined in (3) and (4).

\subsection{Packet-to-Flow State Abstraction}
The transition from packet- to flow-level simulation requires the abstraction from fine-grained per-packet processing to coarse point-to-point flow transmission. At flow-granularity as estimated in (4), the network state is primarily characterized by the allocated bandwidth and the remaining data size to be transmitted of each active flow. Thus, during the packet-to-flow transition, the network states under packet-granularity, including instantaneous bandwidth and cumulative transmitted data size, are abstracted into those under flow-granularity.

\textbf{1) Parameter Synchronization:} According to (2)-(4), accurate flow-level simulation requires synchronizing the key parameters that govern bandwidth allocation and flow progression, for each steady-state flow $f_j \in \mathcal{F}_\mathrm{steady}$, which includes:
\begin{itemize}
    \item \textbf{Reallocated Bandwidth ($B_j^{\rm Re}$):} The instantaneous bandwidth of each active flow $f_j$, measured at the end of latest packet-level simulation, is synchronized to the flow-granularity simulation engine by the bandwidth reallocation process in (2) and (3).
    \item \textbf{Remaining Data Size ($L_j$):} This metric quantifies the residual data to be transmitted in the flow-level simulation. It is derived by subtracting the already-transmitted data size from the total flow size:
    \begin{equation}
        L_j = L_j^{\mathrm{init}} - L_j^{\mathrm{cum}},
    \end{equation}
    where $L_j^{\mathrm{init}}$ represents the initial total size of flow $f_j$, and $L_j^{\mathrm{cum}}$ is the cumulative transmitted data size during previous packet-level simulation. This value is critical for duration estimation of flow-level simulation in (4).
\end{itemize}

\textbf{2) Transmission Time Compensation:} Although (4) estimates the completion time of each steady-state flow, it only captures the transmission time under the reallocated bandwidth and does not account for the queuing delay experienced by in-flight packets at intermediate switches. As a result, the estimated FCT may be underestimated. To improve accuracy, \texttt{HyGra} aggregates the queue occupancy observed along the forwarding path of each flow and converts the resulting queuing effect into an equivalent delay compensation term:
\begin{equation}
    \Delta \tau_j = \sum_{h \in \mathcal{P}_j} \frac{Q_h}{C_h},
\end{equation}
where $\mathcal{P}_j$ denotes the set of intermediate switch hops on the path of flow $f_j$, $Q_h$ is the amount of queued traffic observed at hop $h$, and $C_h$ is the link capacity of hop $h$. The compensated completion time of flow $f_j$ is then given by
\begin{equation}
    \hat{\tau}_j = \tau_j + \Delta \tau_j,
\end{equation}
where $\tau_j$ is the remaining completion time of flow $f_j$ in~(4).

\subsection{Flow-to-Packet State Restoration}
Reverting from flow-granularity back to packet-granularity is more challenging than the packet-to-flow transition. This is because during the flow-level simulation, bandwidth fluctuations induce asymmetric switch queue states at $T_\mathrm{start}$ and $T_\mathrm{end}$. Thus, directly resuming packet-level network state at $T_\mathrm{end}$ via that at $T_\mathrm{start}$ leads to inaccurate FCT estimation. To address this, we introduce a Transformer-based state restoration mechanism. Queue evolution exhibits strong temporal correlation and Transformer’s self-attention mechanism is suited for capturing both short- and long-range dependencies. By learning temporal structure from historical queue observations collected during packet-level simulation, the model infers the queue state at the end of flow-level simulation, enabling near-lossless restoration of network state. This process consists of offline training and online restoration phases:

\textbf{1) Offline Training of the Transformer:} During the offline phase, a set of historical queue-depth traces is sampled to train a Transformer encoder, enabling the model to capture temporal dependencies in switch port dynamics. For each switch port $u$, the queue evolution is represented as a time series:
\begin{equation}
    \mathcal{A}_u^{1:T} = \{a_u^1, a_u^2, \dots, a_u^T\},
\end{equation}
where $a_u^t$ denotes the queue depth observed at time step $t$ of port $u$. Then the observed queue depth sequences are processed in Transformer encoder using the standard query, key, and value formulation:
\begin{equation}
    Q=A_uW^Q, K=A_uW^K, V=A_uW^V
\end{equation}
where $W^{Q}$, $W^{K}$, and $W^{V}$ are learnable projection matrices. These matrices encode the current queue state, historical contextual patterns, and time-step-specific features, respectively. After that, self-attention is utilized to capture the dependencies between time steps\footnote{The formulation of self-attention is given by $\text{Attention(Q, K, V)}= \text{softmax}\left({Q K^{T}}/{\sqrt{d_k}}\right)V$, where $d_k$ is the key dimension\cite{transformer}.}.
\begin{algorithm}[t]
\caption{Transformer-based Queue State Restoration}
\label{algo:queue_reconstruct}
\small
\begin{algorithmic}[1]
\Require Historical queue depth traces $\mathcal{A}_u^{\mathrm{hist}}$;
         Real-time queue sequence per port $\mathcal{A}_u^{1:T_\mathrm{start}}$.
\Ensure Restored steady-state queue depth $\hat{a}_u^{T_\mathrm{end}}$.

\Statex \textbf{\texttt{//Phase 1: Offline Training}}
\State Initialize Transformer model $\psi$ with random weights;
\Repeat
    \State Sample historical sequence $\mathcal{A}_u \in \mathcal{A}_u^{\mathrm{hist}}$;
    \State Compute $Q, K, V$ projections by (10);
    \State Update model parameters to minimize prediction loss;
\Until{model $\psi$ converges}.

\Statex \textbf{\texttt{//Phase 2: Online Restoration}}
\For{each switch port $u$}
    \State Predict state $\hat{a}_u^{T_\mathrm{end}} \leftarrow \psi(\mathcal{A}_u^{1:T_\mathrm{start}})$ by (11);
    \State Update queue state of port $u$ with $\hat{a}_u^{T_\mathrm{end}}$.
\EndFor
\end{algorithmic}
\end{algorithm}

\textbf{2) Online Queue Depth Restoration} \texttt{HyGra} records the queue-depth sequence immediately before switching to the flow-granularity simulation (as described in Algorithm 1). Then, at the end of the flow-granularity simulation, this sequence is provided to the previously-trained Transformer model ($\psi$) to infer the corresponding queue depth:
\begin{equation} 
    \hat{a}_u^{T_\mathrm{end}} = \psi(\mathcal{A}_u^{1:T_\mathrm{start}}). 
\end{equation}
The predicted queue depth $\hat{a}_u^{T_\mathrm{end}}$ is then used to update the corresponding switch-port queue state in the packet-level simulator. Packet-granularity simulation then resumes from the restored state rather than the original state at $T_\mathrm{end}$. The detailed procedure is summarized in Algorithm~2.

\section{Evaluation}
\subsection{Parameter Settings}
\subsubsection{DCN Environment} As illustrated in Fig. \ref{application}\circled{A}, the considered DCN environment employs a Clos topology to interconnect computing nodes. The number of GPU within each node is 8, and the inter-node network links is configured with a bandwidth of 200 Gbps and a propagation latency of 10 µs. As specified in Algorithm~1, we initialize $\epsilon_{bw}=10^{-5}$, $\epsilon_{q}=10^{-4}$, and $N_{\mathrm{stable}}=3$.
\subsubsection{Testbed} \texttt{HyGra} is implemented on SystemC and ns-3, which serve as the foundational platforms for packet-granularity simulation. All experiments are conducted on a Huawei FusionServer 2288H V5 host equipped with dual Intel Xeon Gold 6150 processors, providing 36 physical cores and 72 hardware threads at 2.7 GHz with AVX-512 support, and 384 GiB of six-channel DDR4-2666 ECC memory. The software environment runs Ubuntu 20.04 LTS with the Linux 5.4.0-200-generic kernel.
\subsubsection{Parallelization Strategy} We consider three widely-used parallelization strategies and their mixture:
\begin{itemize}
    \item \textbf{Pipeline Parallelism (PP):} This strategy partitions consecutive model layers to different pipeline stages. Communication is dominated by inter-node transfers of hidden-layer between adjacent stages.
    \item \textbf{Tensor Parallelism (TP):} In this strategy, model parameters are partitioned across multiple GPUs. As a result, communication is dominated by high-frequency synchronization.
    \item \textbf{Data Parallelism (DP):} Different from TP, this strategy replicates the full model across nodes, each processing a distinct subset of the training data. After each iteration, gradients are synchronized globally across all nodes.
    \item \textbf{Mixed:} This case integrates the three parallelization strategies described above.
\end{itemize}

\begin{figure}[t!]
    \centering
    \captionsetup[subfloat]{
        font=scriptsize,
        labelsep=space,
        captionskip=1pt,
        justification=centering
    }
    
    \subfloat[PP]{%
        \includegraphics[width=0.31\linewidth]{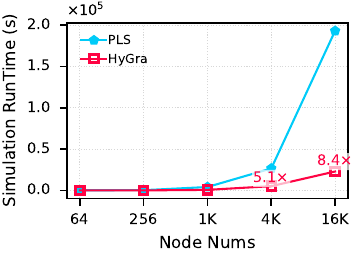}%
        \label{subfig:a}%
    }
    \hfill
    \subfloat[TP]{%
        \includegraphics[width=0.31\linewidth]{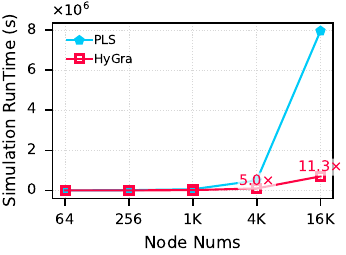}%
        \label{subfig:b}%
    }
    \hfill
    \subfloat[DP]{%
        \includegraphics[width=0.31\linewidth]{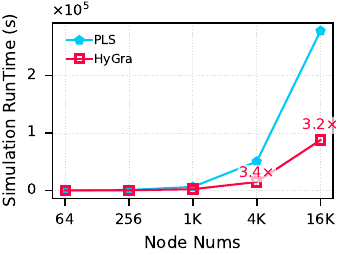}%
        \label{subfig:c}%
    }
    
    \caption{Simulation runtime under individual parallelization strategies for PP, TP, and DP.}
    \label{sim_time}
\end{figure}
\subsubsection{Model Parameters}
We configure three representative commercial LLMs to evaluate performance across diverse model architectures and scales:
\begin{itemize}
    \item \textbf{GPT-3}\cite{gpt3}\textbf{:} This model comprises 175 billion parameters and adopts a standard dense Transformer architecture with 96 layers and a hidden dimension of 12,288. It is trained using the Adam optimizer and a context length of 2,048 tokens.
    \item \textbf{DeepSeek-V3}\cite{deepseekv3}\textbf{:} Scaling to 671 billion parameters, with 37 billion active parameters per token, it adopts a mixture-of-experts (MoE) architecture integrated with multi-head latent attention. Training is performed using AdamW and a peak learning rate of $2.2 \times 10^{-4}$.
    \item \textbf{Qwen2.5}\cite{bai2023qwen}\textbf{:} Comprising 72 billion parameters, this dense model integrates grouped query attention and SwiGLU activations to enhance computational efficiency. It is optimized using AdamW and supports a context window extensible to 128k tokens.
\end{itemize}
\subsubsection{Baselines} We consider two representative simulators as benchmarks to assess the efficiency and fidelity of \texttt{HyGra}:
\begin{itemize}
    \item \textbf{Packet-level Simulator (PLS):} This baseline relies exclusively on the packet-granularity simulation engine (E1 in Fig. \ref{HyGra}) and serves as the upper bound on fidelity.
    \item \textbf{Flow-level Simulator (FLS):} This baseline solely operates on the flow-granularity simulation engine (E2 in Fig. \ref{HyGra}). As discussed in Section II-A, it sacrifices fidelity to maximize execution speed, thus serving as a theoretical upper bound on simulation efficiency in our experiments.
\end{itemize}

\subsubsection{Performance Metrics} As analyzed in Section II-A, we evaluate simulator performance from the following aspects:
\begin{itemize}
    \item \textbf{Efficiency:} We evaluate this by measuring simulator’s computational cost via: (a) \textit{simulation run time}, the wall-clock time required to complete a given task; (b) \textit{CPU utilization}, recording the percentage of processing resources consumed; and (c) \textit{memory footprint}, indicating the amount of RAM allocated during execution.
    \item \textbf{Fidelity:} We quantify this using relative error with respect to the ground truth obtained by the PLS across three metrics: (a) \textit{flow completion time (FCT)}, the time from flow initiation to completion; (b) \textit{job completion time (JCT)}, representing the elapsed time to complete the whole training task; and (c) \textit{throughput}, the data processing rate of inter-node links.
\end{itemize}

\begin{figure}[!t]
    \centering
    \captionsetup[subfloat]{
        font=scriptsize,
        labelsep=space,
        captionskip=1pt,
        justification=centering
    }

    \subfloat[CPU utilization]{%
        \includegraphics[width=0.47\linewidth]{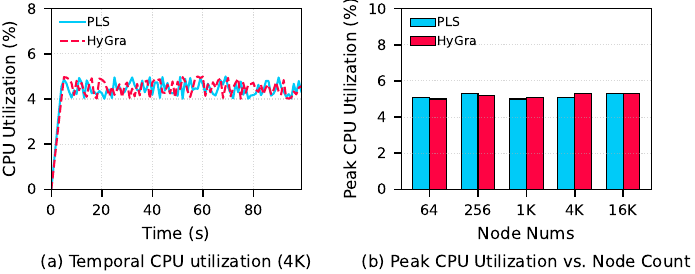}%
        \label{cpu}%
    }
    \hfill
    \subfloat[Memory footprint]{%
        \includegraphics[width=0.47\linewidth]{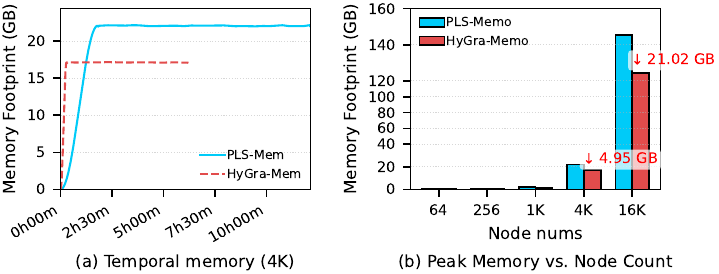}%
        \label{memo}%
    }

    \caption{Resource utilization comparison between different simulators (SystemC, ChatGPT).}
    \label{resource}
\end{figure}
\subsection{Performance Under Individual Parallelization Strategy}
\subsubsection{Efficiency} Fig.~\ref{sim_time} compares the simulation runtime of \texttt{HyGra} and PLS for ChatGPT workloads under different parallelization strategies across varying node scales. As the node count increases, PLS exhibits near-exponential runtime growth under all strategies, whereas \texttt{HyGra} maintains substantially more stable scalability, achieving speedups of 5-8$\times$ under PP, over 11.3$\times$ under TP, and more than 3.3$\times$ under DP. This improvement results from replacing per-packet processing with flow-level simulation during steady phases, reducing unnecessary packet processing. The different speedups across parallelization strategies are mainly determined by the duration and continuity of their steady phases. PP and DP achieve moderate acceleration because their communication phases are relatively short and frequently interrupted. PP is affected by pipeline warm-up and cool-down phases, while DP performs short gradient synchronization operations separated by computation. In contrast, TP achieves the highest acceleration because repeated collective communication among fixed device groups maintains steady phases for longer periods, providing more opportunities for flow-level simulation.

Fig. \ref{cpu} (a) illustrates the temporal evolution of CPU utilization for \texttt{HyGra} and PLS at a node scale of 4k. Both approaches quickly stabilize after initialization and exhibit nearly overlapping curves, fluctuating within a consistently low range, indicating that \texttt{HyGra} introduces negligible additional computational overhead. Fig. \ref{cpu} (b) further reports the peak CPU utilization across node scales ranging from 64 to 16k. In all cases, peak utilization remains below 5\% for both methods, confirming that computational capacity is not the primary bottleneck affecting simulation efficiency.

Besides, Fig. \ref{memo} (a) presents the evolution of memory consumption for \texttt{HyGra} and PLS. \texttt{HyGra} consistently maintains a lower and more stable memory footprint than PLS. Fig. \ref{memo} (b) further compares peak memory usage across node scales. Although both approaches exhibit similar memory demands at small scales, \texttt{HyGra} achieves substantial memory reduction at larger scales, particularly with 4K and 16K nodes. This improved memory efficiency enables large-scale simulation with lower hardware requirements.
\begin{figure}[!t]
    \centering
    \captionsetup[subfloat]{
        font=scriptsize,
        labelsep=space,
        captionskip=1pt,
        justification=centering
    }

    \subfloat[]{%
        \includegraphics[width=0.32\linewidth]{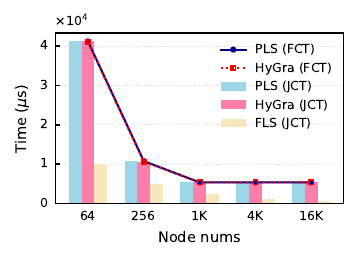}%
        \label{subfig:fidelity_pp}%
    }
    \hfill
    \subfloat[]{%
        \includegraphics[width=0.32\linewidth]{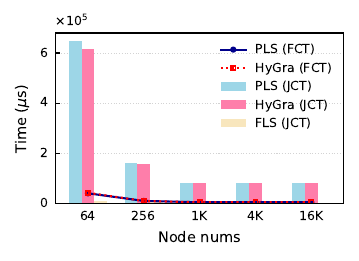}%
        \label{subfig:fidelity_tp}%
    }
    \hfill
    \subfloat[]{%
        \includegraphics[width=0.32\linewidth]{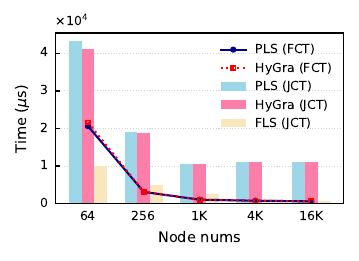}%
        \label{subfig:fidelity_dp}%
    }

    \caption{Job completion time (JCT) and flow completion time (FCT) of different simulators under PP, TP, and DP.}
    \label{fidelity}
\end{figure}
\begin{table}[t!]
  \centering
  \caption{FCT errors (\%) of \texttt{HyGra} related to PLS}
  \label{fct_err}

  \scriptsize
  \setlength{\tabcolsep}{3pt}

  \begin{tabular}{ccccccc}
    \toprule
    \multirow{2}{*}{\textbf{Node Nums}} 
    & \multicolumn{2}{c}{\textbf{PP}} 
    & \multicolumn{2}{c}{\textbf{TP}} 
    & \multicolumn{2}{c}{\textbf{DP}} \\

    \cmidrule(lr){2-3} 
    \cmidrule(lr){4-5} 
    \cmidrule(lr){6-7}

    & Max (\%) & Avg (\%) 
    & Max (\%) & Avg (\%) 
    & Max (\%) & Avg (\%) \\

    \midrule
    64   & 4.54  & 4.54  & 4.51  & 2.38 & 4.46  & 3.34  \\
    256  & 1.54  & 1.54  & 1.53  & 0.82 & 1.48  & 0.87  \\
    1K   & 0.07  & 0.06  & 0.06  & 0.03 & 0.06  & 0.01  \\
    4K   & 0.074 & 0.056 & 0.058 & 0.03 & 0.055 & 0.029 \\
    16K  & 0.07  & 0.06  & 0.03  & 0.03 & 0.03  & 0.03  \\
    \bottomrule
  \end{tabular}

  \vspace{-0.2cm}
\end{table}

\subsubsection{Fidelity} We first evaluate JCT under different parallelization strategies. As shown in Fig. \ref{fidelity}, while FLS suffers from severe accuracy degradation with relative errors exceeding 85\%, \texttt{HyGra} maintains strong fidelity, keeping JCT error below 5\% across all scenarios. Under the PP scenario, the relative error is as low as 1.76\% and further decreases as the node scale increases. Fidelity is further assessed using FCT, as reported in Table \ref{fct_err} and Fig. \ref{fidelity}. The average relative FCT error remains consistently low, ranging from 0.056\% to 4.54\% for PP, 0.03\% to 2.38\% for TP, and 0.029\% to 3.34\% for DP. Importantly, \texttt{HyGra} preserves tail-delay accuracy, with the maximum FCT error remaining below 4.54\%. Table \ref{throughput} further compares throughput simulated by PLS and \texttt{HyGra}. At small scales, the maximum deviation of \texttt{HyGra} from PLS ranges between 1 and 1.5 Gbps, corresponding to approximately 4\% to 5\% relative error. As the scale increases, the discrepancy becomes negligible, with absolute differences around 0.1 Gbps and relative error below 0.1\%. This result demonstrates that \texttt{HyGra} preserves high fidelity.

\begin{table*}[t!]
  \centering
  \caption{Throughput comparison between \texttt{HyGra} \& PLS}
  \label{throughput}
  \footnotesize
  \begin{tabular*}{\textwidth}{@{\extracolsep{\fill}} c ccc ccc ccc @{}}
    \toprule
    \multicolumn{1}{c}{\multirow{2}{*}{\textbf{Node nums}}}
      & \multicolumn{3}{c}{\textbf{PP}} 
      & \multicolumn{3}{c}{\textbf{TP}} 
      & \multicolumn{3}{c}{\textbf{DP}} \\
    \cmidrule(lr){2-4} \cmidrule(lr){5-7} \cmidrule(lr){8-10}
      & PLS (Gbps) & HyGra (Gbps) & Rel.Error  
      & PLS (Gbps) & HyGra (Gbps) & Rel.Error 
      & PLS (Gbps) & HyGra (Gbps) & Rel.Error \\
    \midrule
    64   & 23.72  & 24.85  & 4.76\%  & 24.84  & 23.72  & 4.51\%  & 24.82  & 23.73  & 4.39\% \\
    256  & 94.86  & 96.31  & 1.53\%  & 94.83  & 96.31  & 1.56\%  & 96.10  & 94.67  & 1.49\% \\
    1K   & 189.53 & 189.42 & 0.058\% & 188.38 & 189.48 & 0.058\% & 188.26 & 188.38 & 0.063\% \\
    4K   & 189.56 & 189.42 & 0.074\% & 187.73 & 189.42 & 0.090\% & 187.83 & 187.73 & 0.053\% \\
    \bottomrule
  \end{tabular*}
\end{table*}
\begin{table*}[ht!]
  \centering
  \caption{Simulation performance for mixed parallelization strategies across different platforms (512 nodes).}
  \label{sim_performance_mixed}
  \footnotesize
  \renewcommand{\arraystretch}{1.1} 
  \begin{tabular*}{\textwidth}{@{\extracolsep{\fill}} c c c c c c @{}}
    \toprule
    \multirow{2}{*}{\textbf{PLS Platform}} & \multirow{2}{*}{\textbf{Metric}} & \multicolumn{2}{c}{\textbf{DeepSeek}} & \multicolumn{2}{c}{\textbf{Qwen}} \\ 
    \cmidrule(lr){3-4} \cmidrule(lr){5-6}
    & & PLS & HyGra & PLS & HyGra \\ 
    \midrule
    \multirow{2}{*}{SystemC} 
    & Simulation runtime (s) & 6709 & 3619 (\textbf{1.85$\times$}) & 7358 & 2083 (\textbf{3.53$\times$}) \\
    & FCT error (\%)         & --  & 4.2 (Max) / 3.9 (Avg)       & --  & 4.2 (Max) / 4.09 (Avg) \\
    \midrule
    \multirow{2}{*}{ns-3}    
    & Simulation runtime (s) & 25541 & 1650 (\textbf{15.48$\times$}) & 48487 & 39249 (\textbf{1.24$\times$}) \\
    & FCT error (\%)         & --  & 6.7 (Max) / 4.2 (Avg)        & --  & 2.7 (Max) / 0.9 (Avg) \\
    \bottomrule
  \end{tabular*}
\end{table*}

\subsection{Performance Under Mixed Parallelization Strategies}
We next evaluate the performance of \texttt{HyGra} on DeepSeek and Qwen workloads under mixed parallelization strategies combining PP, TP, DP, and expert parallelism (EP)\cite{deepseekv3-1}. We adopt simulation runtimes and FCT error as representative metrics for efficiency and fidelity, respectively. As shown in Table \ref{sim_performance_mixed}, \texttt{HyGra} achieves a 1.85× speedup on simulation runtime for deepseek and a 3.53× speedup for Qwen while maintaining comparable fidelity. The difference in speedup reflects \texttt{HyGra}'s sensitivity to model-specific communication patterns. DeepSeek’s MoE architecture generates intensive all-to-all traffic that frequently triggers packet-granularity simulation, whereas Qwen’s dense architecture produces more stable flows, thereby extending flow-granularity simulation and maximizing acceleration.

To evaluate the versatility of \texttt{HyGra}, we further assess its performance when using ns-3 as the PLS backend. As shown in Table~\ref{sim_performance_mixed}, \texttt{HyGra} maintains superior performance under ns-3, achieving speedups of 15.48× for DeepSeek and 1.24× for Qwen. The difference in acceleration between SystemC and ns-3 mainly stems from their different packet-level event scheduling mechanisms, with ns-3 adopting a more comprehensive and detailed design than the relatively lightweight SystemC implementation.

\subsection{Necessity of State Restoration}
As discussed in Section V-C, flow-to-packet state restoration is crucial for preventing state loss when reverting from flow-level to packet-level simulation. Table~\ref{discuss_sr} quantifies its impact on fidelity using the maximum per-flow FCT error. Without state restoration, the maximum FCT error reaches 15.32\% and 12.48\% for DeepSeek and Qwen at 128 nodes, increasing to 26.63\% and 16.98\% at 512 nodes. Enabling state restoration significantly reduces these errors to 4.6\% and 1.2\% at 128 nodes, and to 6.7\% and 2.7\% at 512 nodes. This result stems from residual bandwidth fluctuations during steady phases that induce queue-depth variations. As flow-granularity simulation neglects switches and does not capture queue variations, the queue states at entry to and exit from flow-level simulation are generally inconsistent.

\subsection{Comparison With Acceleration Based on Parallel Deployment}
To evaluate the effectiveness of hybrid-granularity based acceleration, table~\ref{single-multi} compares the speedup achieved by the proposed HyGra running on a single CPU core with that of PLS accelerated through parallel execution on multi CPU cores. All speedup values are reported relative to the baseline PLS executed on a single CPU core. It shows that \texttt{HyGra} achieves comparable speedups without multi-core scaling, delivering 16.56$\times$ at 128 nodes versus 15.19$\times$ for multi-core PLS and sustaining 15.48$\times$ at 512 nodes, which remains competitive with 18.17$\times$ under multi-core PLS. These results indicate that \texttt{HyGra} achieves substantial acceleration under limited computational resources without requiring specialized hardware adaptations, thereby facilitating the deployment of network-state simulators on commodity platforms.
\begin{table}[!t]
  \centering
  \caption{Maximum per-flow FCT error (\%) of \texttt{HyGra} with and without state restoration (SR).}
  \label{discuss_sr}
  \footnotesize
  \begin{tabular*}{\columnwidth}{@{\extracolsep{\fill}} c c c c @{}}
    \toprule
    \textbf{Node nums} & \textbf{Configuration} & \textbf{DeepSeek} & \textbf{Qwen} \\
    \midrule
    
    \multirow{2}{*}{128} & w/o SR  & 15.32 & 12.48 \\
                         & with SR & 4.6   & 1.2 \\
                         
    \midrule
    
    \multirow{2}{*}{512} & w/o SR  & 26.63 & 16.98 \\
                         & with SR & 6.7   & 2.7 \\
                         
    \bottomrule
  \end{tabular*}
\end{table}
\begin{table}[!t]
  \centering
  \caption{Speedup of \texttt{HyGra} under single-core and PLS under multi-core (SystemC, DeepSeek).}
  \label{single-multi}
  \footnotesize
  \renewcommand{\arraystretch}{1.2} 
  \begin{tabular*}{\columnwidth}{@{\extracolsep{\fill}} c c c @{}}
    \toprule
    \textbf{Deployment Mode} & \textbf{128 Nodes} & \textbf{512 Nodes} \\
    \midrule
    HyGra (Single-Core) & 16.56$\times$ & 15.48$\times$ \\
    \addlinespace
    PLS (Multi-Core)    & 15.19$\times$ & 18.17$\times$ \\
    \bottomrule
  \end{tabular*}
\end{table}

\section{Potential Applications of HyGra}
Due to the good performance on both efficiency and fidelity, along with easy deployment on a single machine, \texttt{HyGra} is well-suited for optimizing parameter configurations for LLM training in DCNs. Specifically, \texttt{HyGra} supports three key application scenarios:
\begin{itemize}
    \item \textbf{Parallelization Strategy Optimization:} By measuring JCT of LLM training task under different settings of parallelization, the most effective strategy combinations can be identified.
    \item \textbf{Communication Configuration Optimization:} By quantifying communication overhead metrics, including link utilization, queue depth, and FCT during LLM training, topology-aware routing algorithms and congestion control mechanisms in DCNs can be optimized, thus improving communication efficiency and reducing inter-node transmission latency.
    \item \textbf{Computation Resource Optimization:} By analyzing JCT under varying communication configurations and node scales, it is possible to identify the saturation point beyond which adding compute nodes no longer yields substantial JCT improvement, thereby preventing inefficient allocation of computational resources.
\end{itemize}
\begin{figure}[t!]
  \centering
  \includegraphics[width=0.82\linewidth]{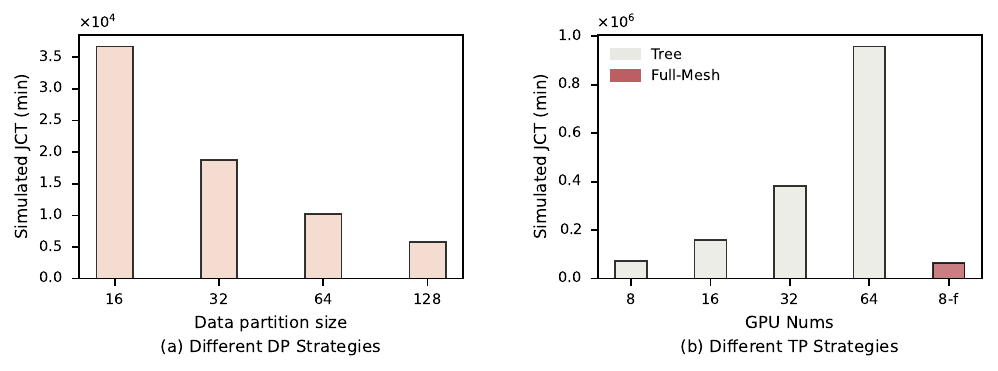}
  \caption{JCT of LLM training under different parallelization strategies (ChatGPT, SystemC).}
  \label{OH}
\end{figure}

To demonstrate this potential, we present two simple case studies to illustrate how \texttt{HyGra} guides training efficiency optimization. The first scenario examines the influence of training parameters under DP strategy. As shown in Fig.~\ref{OH}(a), the simulated JCT increases as the data partition size decreases. This trend arises because smaller partitions require a greater number of iterations to complete a training epoch, thereby increasing the frequency of inter-GPU parameter synchronization. The second one is communication configuration optimization for TP. Fig.~\ref{OH}(b) compares the simulated JCT for completing a TP operation under varying GPU counts and different network topologies, including tree and full-mesh (full connectivity). Under the tree topology, JCT increases significantly with the number of GPUs due to intensified inter-GPU communication. Furthermore, replacing the tree topology with an idealized full-mesh topology under the identical 8-GPU configuration reduces the JCT slightly; however, limited inter-node bandwidth remains the dominant bottleneck constraining TP efficiency. Above case studies demonstrate that \texttt{HyGra} accurately captures communication performance across diverse training configurations, establishing it as an effective tool for optimizing LLM training efficiency in DCNs.

It is worth noting that, as shown in our evaluation in Section V, \texttt{HyGra} introduces only small relative errors in FCT, JCT, and throughput, while Fig.~\ref{OH} shows substantial performance differences among the evaluated configurations. Therefore, these errors are unlikely to change the ranking of configurations with clear performance gaps, enabling \texttt{HyGra} to identify better-performing configurations during design-space exploration. For special cases where the performance differences among candidate configurations evaluated by \texttt{HyGra} are insignificant, PLS can be further applied to identify the best configuration. This evaluation process preserves the efficiency of large-scale exploration while improving the reliability of final decisions.

\section{Conclusion}
In this paper, we propose \texttt{HyGra}, a single-machine hybrid-granularity network-state simulator for LLM training over DCNs. Unlike existing approaches that operate solely at packet or flow granularity and struggle to balance efficiency and fidelity, \texttt{HyGra} exploits the intrinsic network dynamics of LLM training to adaptively switch between granularities. Specifically, \texttt{HyGra} adopts packet-level simulation during non-steady phases characterized by transient fluctuations, and flow-level simulation during steady phases with periodic communication patterns. It further enables near-lossless transitions between these granularities, thereby significantly accelerating execution while preserving high simulation fidelity. Moreover, \texttt{HyGra} requires no specialized hardware, can be readily deployed on a single machine, and remains compatible with established PLSs. Extensive experiments based on representative commercial LLM workloads, including ChatGPT, DeepSeek, and Qwen, demonstrate that \texttt{HyGra} achieves up to 15.4× speedup under single parallelization strategies and 7.8× under mixed parallelization strategies.

We further investigate the applicability of \texttt{HyGra} to parallelization strategy optimization, communication configuration tuning, and computational resource allocation. Preliminary case studies demonstrate that \texttt{HyGra} accurately captures communication performance across diverse training configurations, establishing it as an effective and practical tool for optimizing LLM training efficiency in DCNs. Nevertheless, its acceleration depends on the duration of steady-state phases and may diminish under rapidly changing traffic or when fine-grained packet-level dynamics dominate the target metrics. In addition, the current implementation focuses primarily on network-state simulation and does not yet model the detailed interaction between communication and computation. Future work will explore easily deployable parallel execution across multiple CPU cores to further improve simulation speedup, and incorporate computing-state simulation to enable network-compute co-simulation for LLM training in DCNs.

\section*{Acknowledgment}
We thank Lu Zhao and Peijie Sun for their feedback on this paper.

\bibliographystyle{IEEEtran}
\bibliography{refer}

@techreport{gpt4,
  author    = {{OpenAI}},
  title     = {{Gpt-4} technical report},
  year      = {2024},
  institution = {OpenAI}
}

@article{llama3,
  title={The {llama} 3 herd of models},
  author={Grattafiori, Aaron and Dubey, Abhimanyu and Jauhri, Abhinav and Pandey, Abhinav and Kadian, Abhishek and Al-Dahle, Ahmad and Letman, Aiesha and Mathur, Akhil and Schelten, Alan and Vaughan, Alex and others},
  journal={arXiv preprint arXiv:2407.21783},
  year={2024}
}

@inproceedings{sc21,
  title={Efficient large-scale language model training on gpu clusters using {Megatron-LM}},
  author={Narayanan, Deepak and Shoeybi, Mohammad and Casper, Jared and LeGresley, Patrick and Patwary, Mostofa and Korthikanti, Vijay and Vainbrand, Dmitri and Kashinkunti, Prethvi and Bernauer, Julie and Catanzaro, Bryan and others},
  booktitle={International conference for high performance computing, networking, storage and analysis (SC)},
  pages={1--15},
  year={2021}
}

@inproceedings{sc25,
    author = {P\'{e}rez Di\'{e}guez, Adri\'{a}n and Batlle Casellas, \`{A}lex and Torres-Camps, Aleix and Teague, Harris and Ros-Giralt, Jordi},
    title = {Pretraining LLMs at Scale: Tuning Strategies and Performance Portability.},
    year = {2025},
    doi = {10.1145/3731599.3767699},
    pages = {1512–1523},
    numpages = {12},
    series = {SC Workshops '25}
}

@inproceedings{2010dctcp,
  title={Data center tcp (dctcp)},
  author={Alizadeh, Mohammad and Greenberg, Albert and Maltz, David A and Padhye, Jitendra and Patel, Parveen and Prabhakar, Balaji and Sengupta, Sudipta and Sridharan, Murari},
  booktitle={ACM SIGCOMM},
  pages={63--74},
  year={2010}
}

@inproceedings{2025spme,
  title={Accelerating Design Space Exploration for {LLM} Training Systems with Multi-experiment Parallel Simulation},
  author={Gui, Fei and Gao, Kaihui and Chen, Li and Li, Dan and Liu, Vincent and Zhang, Ran and Yang, Hongbing and Xiong, Dian},
  booktitle={USENIX Symposium on Networked Systems Design and Implementation (NSDI)},
  pages={473--488},
  year={2025}
}

@article{2004pdns,
  title={A federated approach to distributed network simulation},
  author={Riley, George F and Ammar, Mostafa H and Fujimoto, Richard M and Park, Alfred and Perumalla, Kalyan and Xu, Donghua},
  journal={ACM Transactions on Modeling and Computer Simulation (TOMACS)},
  volume={14},
  number={2},
  pages={116--148},
  year={2004},
}

@inproceedings{osdi22,
  title={Alpa: Automating inter-and Intra-Operator parallelism for distributed deep learning},
  author={Zheng, Lianmin and Li, Zhuohan and Zhang, Hao and Zhuang, Yonghao and Chen, Zhifeng and Huang, Yanping and Wang, Yida and Xu, Yuanzhong and Zhuo, Danyang and Xing, Eric P and others},
  booktitle={USENIX Symposium on Operating Systems Design and Implementation (OSDI)},
  pages={559--578},
  year={2022}
}

@inproceedings{sigcomm24,
  title={Alibaba {HPN}: A data center network for large language model training},
  author={Qian, Kun and Xi, Yongqing and Cao, Jiamin and Gao, Jiaqi and Xu, Yichi and Guan, Yu and Fu, Binzhang and Shi, Xuemei and Zhu, Fangbo and Miao, Rui and others},
  booktitle={ACM SIGCOMM},
  pages={691--706},
  year={2024}
}

@inproceedings{nsdi23,
  title={{TopoOpt}: Co-optimizing network topology and parallelization strategy for distributed training jobs},
  author={Wang, Weiyang and Khazraee, Moein and Zhong, Zhizhen and Ghobadi, Manya and Jia, Zhihao and Mudigere, Dheevatsa and Zhang, Ying and Kewitsch, Anthony},
  booktitle={USENIX Symposium on Networked Systems Design and Implementation (NSDI)},
  pages={739--767},
  year={2023}
}

@inproceedings{2025simai,
  title={{SimAI}: Unifying Architecture Design and Performance Tuning for {Large-Scale} Large Language Model Training with Scalability and Precision},
  author={Wang, Xizheng and Li, Qingxu and Xu, Yichi and Lu, Gang and Li, Dan and Chen, Li and Zhou, Heyang and Zheng, Linkang and Zhang, Sen and Zhu, Yikai and others},
  booktitle={USENIX Symposium on Networked Systems Design and Implementation (NSDI)},
  pages={541--558},
  year={2025}
}

@inproceedings{li2024m3,
  title={m3: Accurate flow-level performance estimation using machine learning},
  author={Li, Chenning and Nasr-Esfahany, Arash and Zhao, Kevin and Noorbakhsh, Kimia and Goyal, Prateesh and Alizadeh, Mohammad and Anderson, Thomas E},
  booktitle={ACM SIGCOMM},
  pages={813--827},
  year={2024}
}

@inproceedings{simgrid,
  title={Simgrid: A generic framework for large-scale distributed experiments},
  author={Casanova, Henri and Legrand, Arnaud and Quinson, Martin},
  booktitle={Tenth International Conference on Computer Modeling and Simulation (uksim)},
  pages={126--131},
  year={2008},
  organization={IEEE}
}

@misc{ns3,
  author       = {Riley, George F. and Henderson, Thomas R.},
  title        = {The ns-3 Network Simulator},
  year         = {2010},
  howpublished = {Chapter in: Modeling and Tools for Network Simulation, edited by Klaus Wehrle, Mesut G{\"u}ne{\c{s}}, and James Gross, Springer Berlin Heidelberg},
  pages        = {15--34},
  doi          = {10.1007/978-3-642-12331-3_2},
  note         = {Originally published as Chapter 2 of the mentioned book.}
}

@incollection{OMNET++,
  title={A practical introduction to the {OMNeT++} simulation framework},
  author={Varga, Andras},
  booktitle={Recent advances in network simulation: the OMNeT++ environment and its ecosystem},
  pages={3--51},
  year={2019},
  publisher={Springer}
}

@article{binkert2011gem5,
  title={The gem5 simulator},
  author={Binkert, Nathan and Beckmann, Bradford and Black, Gabriel and Reinhardt, Steven K and Saidi, Ali and Basu, Arkaprava and Hestness, Joel and Hower, Derek R and Krishna, Tushar and Sardashti, Somayeh and others},
  journal={ACM SIGARCH computer architecture news},
  volume={39},
  number={2},
  pages={1--7},
  year={2011},
  publisher={ACM New York, NY, USA}
}

@article{deepseekv3,
  title={Deepseek-v3 technical report},
  author={Liu, Aixin and Feng, Bei and Xue, Bing and Wang, Bingxuan and Wu, Bochao and Lu, Chengda and Zhao, Chenggang and Deng, Chengqi and Zhang, Chenyu and Ruan, Chong and others},
  journal={arXiv preprint arXiv:2412.19437},
  year={2024}
}

@inproceedings{2024megascale,
  title={{MegaScale}: Scaling large language model training to more than 10,000 GPUs},
  author={Jiang, Ziheng and Lin, Haibin and Zhong, Yinmin and Huang, Qi and Chen, Yangrui and Zhang, Zhi and Peng, Yanghua and Li, Xiang and Xie, Cong and Nong, Shibiao and others},
  booktitle={USENIX Symposium on Networked Systems Design and Implementation (NSDI)},
  pages={745--760},
  year={2024}
}

@inproceedings{2024crux,
  title={Crux: Gpu-efficient communication scheduling for deep learning training},
  author={Cao, Jiamin and Guan, Yu and Qian, Kun and Gao, Jiaqi and Xiao, Wencong and Dong, Jianbo and Fu, Binzhang and Cai, Dennis and Zhai, Ennan},
  booktitle={ACM SIGCOMM},
  pages={1--15},
  year={2024}
}

@inproceedings{2025atlahs,
  title={Atlahs: An application-centric network simulator toolchain for ai, hpc, and distributed storage},
  author={Shen, Siyuan and Bonato, Tommaso and Hu, Zhiyi and Jordan, Pasquale and Chen, Tiancheng and Hoefler, Torsten},
  booktitle={International Conference for High Performance Computing, Networking, Storage and Analysis (SC)},
  pages={349--367},
  year={2025}
}

@inproceedings{2020DaydreamAE,
  title={Daydream: Accurately estimating the efficacy of optimizations for {DNN} training},
  author={Zhu, Hongyu and Phanishayee, Amar and Pekhimenko, Gennady},
  booktitle={USENIX Annual Technical Conference (ATC)},
  pages={337--352},
  year={2020}
}

@inproceedings{unison,
  title={Unison: a parallel-efficient and user-transparent network simulation kernel},
  author={Bai, Songyuan and Zheng, Hao and Tian, Chen and Wang, Xiaoliang and Liu, Chang and Jin, Xin and Xiao, Fu and Xiang, Qiao and Dou, Wanchun and Chen, Guihai},
  booktitle={European Conference on Computer Systems (EuroSys)},
  pages={115--131},
  year={2024}
}

@inproceedings{dons,
  title={Dons: Fast and affordable discrete event network simulation with automatic parallelization},
  author={Gao, Kaihui and Chen, Li and Li, Dan and Liu, Vincent and Wang, Xizheng and Zhang, Ran and Lu, Lu},
  booktitle={ACM SIGCOMM},
  pages={167--181},
  year={2023}
}

@inproceedings{simbricks,
  title={SimBricks: end-to-end network system evaluation with modular simulation},
  author={Li, Hejing and Li, Jialin and Kaufmann, Antoine},
  booktitle={ACM SIGCOMM},
  pages={380--396},
  year={2022}
}

@book{max-min,
  title={Data networks},
  author={Bertsekas, Dimitri and Gallager, Robert},
  year={2021},
  publisher={Athena Scientific}
}

@inproceedings{deepqueuenet,
  title={DeepQueueNet: Towards scalable and generalized network performance estimation with packet-level visibility},
  author={Yang, Qingqing and Peng, Xi and Chen, Li and Liu, Libin and Zhang, Jingze and Xu, Hong and Li, Baochun and Zhang, Gong},
  booktitle={ACM SIGCOMM},
  pages={441--457},
  year={2022}
}

@article{bai2023qwen,
  title={Qwen technical report},
  author={Bai, Jinze and Bai, Shuai and Chu, Yunfei and Cui, Zeyu and Dang, Kai and Deng, Xiaodong and Fan, Yang and Ge, Wenbin and Han, Yu and Huang, Fei and others},
  journal={arXiv preprint arXiv:2309.16609},
  year={2023}
}

@inproceedings{2024CL,
  title={Evaluating Chiplet-based Large-Scale Interconnection Networks via Cycle-Accurate Packet-Parallel Simulation},
  author={Yinxiao Feng and Yuchen Wei and Dong Xiang and Kaisheng Ma},
  booktitle={USENIX Annual Technical Conference},
  year={2024}
}

@inproceedings{2023MF,
  title={Solving {Max-Min} Fair Resource Allocations Quickly on Large Graphs},
  author={Namyar, Pooria and Arzani, Behnaz and Kandula, Srikanth and Segarra, Santiago and Crankshaw, Daniel and Krishnaswamy, Umesh and Govindan, Ramesh and Raj, Himanshu},
  booktitle={USENIX Symposium on Networked Systems Design and Implementation (NSDI)},
  pages={1937--1958},
  year={2024}
}

@inproceedings{deepseekv3-1,
  title={Insights into deepseek-v3: Scaling challenges and reflections on hardware for ai architectures},
  author={Zhao, Chenggang and Deng, Chengqi and Ruan, Chong and Dai, Damai and Gao, Huazuo and Li, Jiashi and Zhang, Liyue and Huang, Panpan and Zhou, Shangyan and Ma, Shirong and others},
  booktitle={Annual International Symposium on Computer Architecture},
  pages={1731--1745},
  year={2025}
}

@article{li2024splitsim,
  title={SplitSim: Large-Scale Simulations for Evaluating Network Systems Research},
  author={Li, Hejing and Balasubramanian, Praneeth and Meiers, Marvin and Li, Jialin and Kaufmann, Antoine},
  journal={arXiv preprint arXiv:2402.05312},
  year={2024}
}

@inproceedings{pdes,
  title={Parallel discrete event simulation: The making of a field},
  author={Fujimoto, Richard M and Bagrodia, Rajive and Bryant, Randal E and Chandy, K Mani and Jefferson, David and Misra, Jayadev and Nicol, David and Unger, Brian},
  booktitle={Winter Simulation Conference (WSC)},
  pages={262--291},
  year={2017}
}

@inproceedings{LLMC2,
  title={Network load balancing with parallel flowlets for AI training clusters},
  author={Cao, Peirui and Cheng, Wenxue and Zhao, Shizhen and Xiong, Yongqiang},
  booktitle={SIGCOMM Workshop on Networks for AI Computing},
  pages={18--25},
  year={2024}
}

@article{m4,
  title={m4: A Learned Flow-level Network Simulator},
  author={Li, Chenning and Zabreyko, Anton A and Nasr-Esfahany, Arash and Zhao, Kevin and Goyal, Prateesh and Alizadeh, Mohammad and Anderson, Thomas},
  journal={arXiv preprint arXiv:2503.01770},
  year={2025}
}

@article{DP,
  title={ZeRO-Infinity: Breaking the GPU Memory Wall for Extreme Scale Deep learning},
  author={Samyam Rajbhandari and Olatunji Ruwase and Jeff Rasley and Shaden Smith and Yuxiong He},
  journal={International Conference for High Performance Computing, Networking, Storage and Analysis (SC)},
  year={2021},
  pages={1-15}
}

@article{PP,
  title={Advances of Pipeline Model Parallelism for Deep Learning Training: An Overview},
  author={Guan, Lei and Li, Dong-Sheng and Liang, Ji-Ye and Wang, Wen-Jian and Ge, Ke-Shi and Lu, Xi-Cheng},
  journal={Journal of Computer Science and Technology},
  volume={39},
  number={3},
  pages={567--584},
  year={2024}
}

@article{gpt3,
  title={Language Models are Few-Shot Learners},
  author={Brown, Tom B and Mann, Benjamin and Ryder, Nick and Subbiah, Melanie and Kaplan, Jared and Dhariwal, Prafulla and Neelakantan, Arvind and Shyam, Pranav and Sastry, Girish and Askell, Amanda and others},
  journal={Advances in Neural Information Processing Systems},
  volume={33},
  year={2020}
}

@article{transformer,
  title={Attention is all you need},
  author={Vaswani, Ashish and Shazeer, Noam and Parmar, Niki and Uszkoreit, Jakob and Jones, Llion and Gomez, Aidan N and Kaiser, {\L}ukasz and Polosukhin, Illia},
  journal={Advances in neural information processing systems},
  volume={30},
  year={2017}
}

@inproceedings{mmf-wf,
  title={Upward max min fairness},
  author={Danna, Emilie and Hassidim, Avinatan and Kaplan, Haim and Kumar, Alok and Mansour, Yishay and Raz, Danny and Segalov, Michal},
  booktitle={INFOCOM},
  pages={837--845},
  year={2012},
}

@inproceedings{wormhole,
  title={Supercharging Packet-level Network Simulation of Large Model Training via Memoization and Fast-Forwarding},
  author={Fei Long and Kaihui Gao and Li Chen and Dan Li and Yiwei Zhang and Fei Gui and Yitao Xing and Wenjia Wei and Bingyan Liu},
  year={2026}
}

\vfill

\end{document}